\documentclass[
]{ceurart}

\usepackage{listings}
\usepackage{graphicx}
\usepackage{subcaption}

\begin{document}




\copyrightyear{2024}
\copyrightclause{Copyright for this paper by its authors.
  Use permitted under Creative Commons License Attribution 4.0
  International (CC BY 4.0).}

\conference{SIGIR eCom '24: The SIGIR Workshop on eCommerce, 2024, Washington, DC, USA}

\title{Shopping Queries Image Dataset (SQID): An Image-Enriched ESCI Dataset for Exploring Multimodal Learning in Product Search}

\author[1]{Marie Al Ghossein}[%
email=marie.alghossein@crossingminds.com
]
\fnmark[1]
\address[1]{Crossing Minds, San Francisco, CA, USA}

\author[1]{Ching-Wei Chen}[%
email=chingwei.chen@crossingminds.com
]
\fnmark[1]

\author[2]{Jason Tang}[%
]
\fnmark[2]
\address[2]{Stripe, Toronto, ON, Canada}

\fntext[1]{These authors contributed equally.}
\fntext[2]{Work done while at Crossing Minds}
\begin{abstract}
Recent advances in the fields of Information Retrieval and Machine Learning have focused on improving the performance of search engines to enhance the user experience, especially in the world of online shopping. The focus has thus been on leveraging cutting-edge learning techniques and relying on large enriched datasets. This paper introduces the Shopping Queries Image Dataset (SQID), an extension of the Amazon Shopping Queries Dataset enriched with image information associated with 190,000 products. By integrating visual information, SQID facilitates research around multimodal learning techniques that can take into account both textual and visual information for improving product search and ranking. We also provide experimental results leveraging SQID and pretrained models, showing the value of using multimodal data for search and ranking. SQID is available under: https://github.com/Crossing-Minds/shopping-queries-image-dataset.
\end{abstract}


\maketitle

\section{Introduction}
In the age of online shopping, eCommerce platforms must help customers find what they are looking for with the least amount of effort. Product search allows users to enter a search query, and get back a list of results matching that query. An effective product search should be able to understand exactly what a user is looking for, and retrieve the most relevant results from a catalog of available items. To effectively fulfill a user's shopping needs, a search engine must draw on all the information it has available, including textual, visual, and contextual metadata associated with the user, the search query, and the products in the catalog. 

In particular, visual information can be very useful to identify characteristics of products that may not be well represented in textual metadata. To illustrate this point, consider the product listing for a men's dress shirt\footnote{\url{https://www.amazon.com/dp/B07C95ZCP2}}, which includes textual metadata such as:

\begin{itemize}
    \item \textbf{Title}: ``Men's Vertical Striped Slim Fit Long Sleeve Dress Shirt''
    \item \textbf{Description}: ``This Stylish Men's Collared Dress Shirt Comes in a Modern Fit Which is Slightly More of a Tailored Fit Than a Regular Fit. It Also Features Slim Fit, Vertical Striped Printed Pattern, Buttoned Up Closure, Turn Down Collar, Single Breasted Buttons, Convertible Double French Cuff, Round Curved Shirttail Hem''
    \item \textbf{Size Options}: Small, Medium, Large, Extra Large
    \item \textbf{Color Options}: Black, Blue, Navy Blue White, Khaki, White Stripe Black, White/Purple Stripe, Grey Plaid
\end{itemize}

\begin{figure}[htbp]
    \centering
    \begin{subfigure}[b]{0.32\textwidth}
        \centering
        \includegraphics[width=0.6\textwidth]{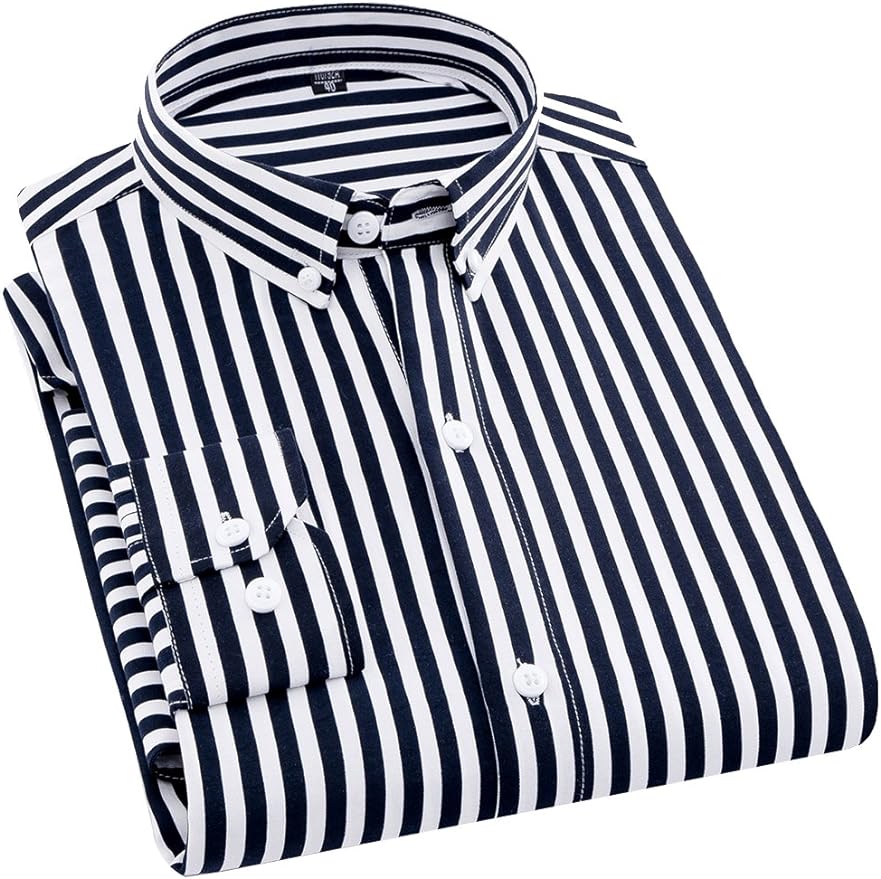}
        \caption{A striped shirt}
        \label{fig:image1}
    \end{subfigure}
    \begin{subfigure}[b]{0.32\textwidth}
        \centering
        \includegraphics[width=0.6\textwidth]{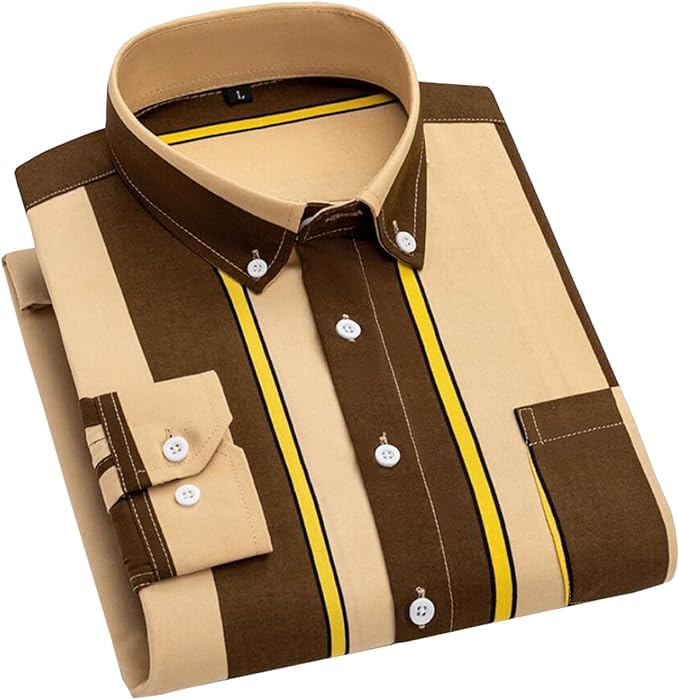}
        \caption{Same shirt, color ``Khaki''}
        \label{fig:image2}
    \end{subfigure}
    \begin{subfigure}[b]{0.32\textwidth}
        \centering
        \includegraphics[width=0.5\textwidth]{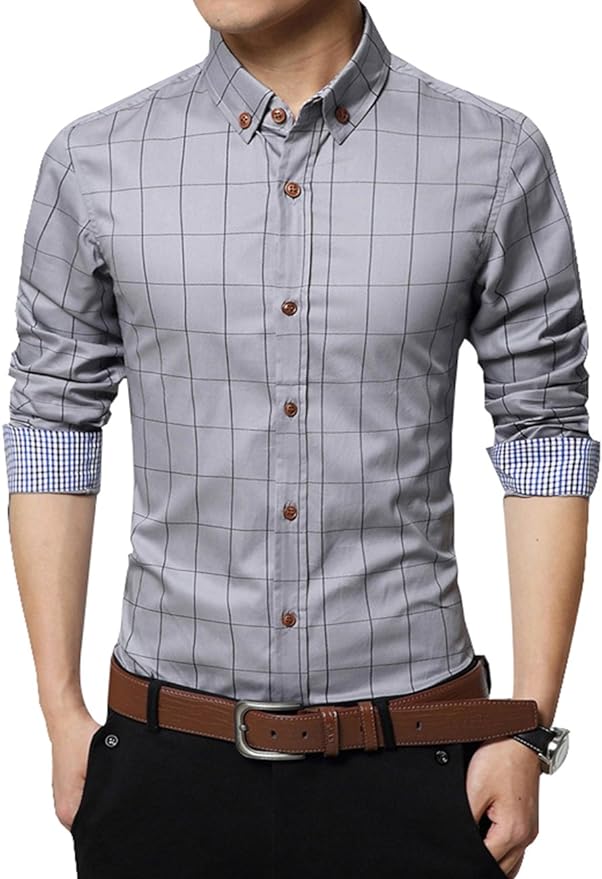}
        \caption{Same shirt, color ``Gray Plaid''}
        \label{fig:image3}
    \end{subfigure}
    \caption{Different variants of a product on Amazon}
    \label{fig:shirts}
\end{figure}

If a user is looking for a ``men's dress shirt with thin vertical stripes'', they might expect that this product is a relevant match based solely on the textual metadata. However, when looking at the product images, they would quickly notice that the stripe pattern on the shirt is not ``thin'' but rather ``thick'' stripes. Not only that, but many of the different color options in fact have a completely different design and thickness of stripes, while some color options have a checkered pattern instead of stripes (Figure~\ref{fig:shirts}).  

None of these options would be a great match for the ``thin vertical stripes'' the user is looking for. However, search engines that rely only on textual metadata are likely to return these shirts as relevant results. If, on the other hand, the search engine leveraged multimodal information such as the product images, it might not have made that mistake.

In order to support research on improving product search by leveraging image information, we are releasing the Shopping Queries Image Dataset (SQID) - an augmented version of the Amazon Shopping Queries Dataset (SQD)\footnote{\url{https://github.com/amazon-science/esci-data}}~\cite{esci} that includes image information and visual embeddings for over 190,000 products, as well as text embeddings of associated search queries so that researchers can explore the effects of multimodal learning on the effectiveness of product search. The dataset is available under: https://github.com/Crossing-Minds/shopping-queries-image-dataset.

The paper is structured as follows. Section~\ref{sec:related_work} presents related work around SQD and pretrained models used to embed multimodal data. Section~\ref{sec:sqid} provides the details of the data covered in SQID, as well as the methodology followed for data collection. Section~\ref{sec:eval} presents the experimental setting used in this paper to highlight the benefit of using multimodal data for ranking, followed by the experimental results provided in Section~\ref{sec:exp_results}.

\section{Related Work} \label{sec:related_work}

This section covers work related to SQD based on which SQID is built, on one hand, and multimodal learning techniques allowing to leverage image and text data for representing items and products, on the other hand. 

\subsection{Shopping Queries Dataset (SQD)}
In 2022, Amazon released the Shopping Queries Dataset (SQD)~\cite{esci}, as part of the KDD Cup challenge. This dataset includes a large number of product search queries from real Amazon users, along with a list of up to 40 potentially relevant results for each query. Each of these results comes with a judgment of how relevant the product is to the search query. These judgments (E, S, C, and I) are described on the KDD Cup'22 Challenge Page\footnote{\url{https://www.aicrowd.com/challenges/esci-challenge-for-improving-product-search}} and correspond to Exact (E), Substitute (S), Complement (C), and Irrelevant (I) (see more details in Table~\ref{escilabels}).


\begin{table*}[th]
\caption{ESCI Label Definitions}
\label{escilabels}
\begin{center}
\vspace{-1.5em}
\begin{tabular}{| c | c | c | c |}
\hline
\bf{ESCI Label} & \bf{Definition} & \bf{Query} & \bf{Product} \\
\hline\hline
Exact (E) & Satisfies all constraints & 
    Thin Vertical Stripes Shirt & Thin Vertical Stripes Shirt \\
Substitute (S) & Alternative substitute &
    Thin Vertical Stripes Shirt & Khaki Shirt \\
Complement (C) & Complements the desired item & 
    Thin Vertical Stripes Shirt & Grey Jacket \\
Irrelevant (I) & Everything else & 
    Thin Vertical Stripes Shirt & Wooden Chair \\
\hline
\end{tabular}
\end{center}
\vskip -0.2in
\end{table*}

The dataset was released along with three tasks\footnote{\url{https://github.com/amazon-science/esci-data?tab=readme-ov-file\#introduction}}:

\begin{itemize}
    \item Task 1 - Query-Product Ranking: Given a query and a set of retrieved products for this query, the goal is to rank the products going from the most relevant to the least relevant, similar to the output of a search engine. 
    \item Task 2 - Multi-class Product Classification: Given a query and a set of retrieved products for this query, the goal is to classify each product as part of the E, S, C, and I classes of products.
    \item Task 3 - Product Substitute Identification: Given a query and a set of retrieved products for this query, the goal is to identify the substitute products from the list of retrieved products.
\end{itemize}

In the context of this challenge, a variety of techniques were explored to improve the score for each of the tasks, including self-distillation, data augmentation, and adversarial training, among others~\cite{lin2022winning, solution2, solution3, wu2022some}.

SQD has also been used to support other use cases. For instance, Tang et al.~\cite{jason_wsdm} generate textual product descriptions based on product images, use it to improve search and recommendation, and evaluate their approach on the Task 1 of the ESCI dataset. Hou et al.~\cite{blair} introduce a set of pretrained sentence embedding models for recommendation and trained on the ``Amazon Reviews 2023'' dataset, a dataset including user reviews and item metadata from Amazon. The ESCI dataset is used to evaluate the performance of these models for conventional product search. 

On another note, the TREC Product Search Track of 2023~\cite{trec2023} leveraged SQD to create a benchmark of retrieval methods used for product search. The dataset was enriched with multimodal data and additional evaluation queries and labels, to make it more suitable for an end-to-end retrieval benchmark rather than a ranking task. Compared to this work, our focus is more aligned with the initial ranking task of the KDD Cup'22. We also document the details of data collection and release textual and visual embeddings as well as experimental results comparable to the ESCI benchmark~\cite{esci}.

\subsection{Multimodal Pretrained Models}

Multimodal pretrained models emerged as a powerful paradigm allowing to learn joint representations capturing the relationships between different types of data such as images, text, and audio. In particular, Contrastive Language-Image Pre-training (CLIP)~\cite{clip} relies on a transformer-based architecture, is trained using a contrastive learning approach, and learns to associate images with corresponding textual data by maximizing the similarity between image-text pairs and minimizing it otherwise. 

In this paper, we rely on CLIP to embed queries and products based on text and image data. While fine-tuning pretrained models on a dataset specific to the task is very beneficial to improve the performance, we consider it outside of the scope of this paper and only use pretrained models in our experiments (more details in Section~\ref{sec:exp_results}).

\section{Shopping Queries Image (SQID) Dataset} \label{sec:sqid}

\subsection{Data Characteristics} \label{subsec:data_charac}
The Shopping Queries Image Dataset (SQID) builds upon SQD by including image information and visual embeddings for each product, as well as text embeddings for the associated queries which can be used for baseline product ranking benchmarking. The image information can be used to enhance or improve the accuracy of product search algorithms by allowing them to leverage multimodal machine learning techniques. 

The image information included in this dataset includes:
\begin{enumerate}
    \item Image URL
    \item Image Embeddings extracted using a CLIP model~\cite{clip}, specifically clip-vit-large-patch14\footnote{\url{https://huggingface.co/openai/clip-vit-large-patch14}}
\end{enumerate}

The original SQD includes two subsets of data: a reduced set (``small\_version'' = 1), used for Task 1 Query-Product Ranking, and a larger set (``large\_version'' = 1), used for Tasks 2 and 3. The queries are also from 3 different locales: ``us'', ``es'', and ``jp''. We limited the scope of this dataset to the following subset of SQD:
\begin{itemize}
    \item ``small\_version'' = 1 (reduced set)
    \item ``product\_locale'' = ``us''
\end{itemize}

The reduced set consists of 1,118,011 <query, rating> judgements, out of which 601,354 are from locale ``us''. These judgments contain references to 482,105 unique products (with a unique product\_id).

We then mainly focus on the products found in the test set of SQD's Task 1 (i.e., having ``split'' = ``test''). The total number of products appearing there is 181,701, out of which 164,900 are unique. While the rest of the paper focuses on this set of products, SQID also includes supplementary data, covering additional products appearing in at least 2 query judgements in the English subset of Task 1. There are 27,139 unique products meeting this criteria and that are not in the test split. 

Overall, therefore, SQID covers 164,900 products, with a supplementary part covering an additional set of 27,139 products. 

\subsection{Data Collection}

\textbf{Image URLs.}
We scraped the Amazon website to retrieve the URL to the main product image displayed on the product page of 164,900 products, resulting in 156,545 product\_id's having an image URL (95\%). We focused on the following domains, attempting to retrieve product pages from each of these successively: .com, .ca, .com.au, .cn, .fr, .de, and .co.jp. There are two main cases for when a product does not have an image URL: 
\begin{itemize}
    \item The product\_id failed to return a valid product page, usually when the product is no longer offered on Amazon, or
    \item There was no image associated with the product - to be precise, the main image of the product is a blank image that says ``No image available''.
\end{itemize}
There are 442 products where the image URL contains this particular URL: \textit{https://m.media-amazon.com/images/G/01/digital/video/web/Default\_Background\_Art\_LTR.\_SX1080\_FMjp\_.jpg}. These are ``generic'' product images for digital video products where there is no product-specific image.
\\
\\
\textbf{Textual and visual embeddings.} In addition to product image URLs, SQID also includes visual and textual embeddings of products. These were obtained using CLIP pretrained model, specifically clip-vit-large-patch14\footnote{\url{https://huggingface.co/openai/clip-vit-large-patch14}}, and based on product image URLs and product titles. To address the product ranking task, we also include query embeddings obtained based on the query text. 

\section{Evaluation} \label{sec:eval}

In order to illustrate the value of using multimodal data for product ranking, we leverage SQID for the Task 1 of the KDD Cup 2022 consisting of query-product ranking~\cite{esci}.

We evaluate the performance of several ranking approaches for the Task 1 (``small\_version'' = 1) on the test set (``split'' = ``test'') and for the US locale (``product\_locale'' = ``us''). The evaluation dataset consists of 181,701 judgements, 8,956 queries, and 164,900 products. The average number of judgements per query is around 20.

We only rely on pretrained models and consider that fine-tuning models on the ESCI training data as well as other more advanced techniques used by winning solutions of the challenge (e.g.,~\cite{lin2022winning}) are outside the scope of this paper and to be investigated in future work. 

\subsection{Metrics} Following the setting of the challenge, the ranking quality is measured using the Normalized Discounted Cumulative Gain (NDCG)~\cite{esci}. The four degrees of relevance of a product to a query, defined by the labels E (Exact), S (Substitute), C (Complement), and I (Irrelevant), are attributed respectively to the following relevance scores: 1.0, 0.1, 0.01, and 0.0. To ensure reproducibility and follow the same guidelines as the ESCI benchmark\footnote{\url{https://github.com/amazon-science/esci-data/}}, we use the Terrier IR platform\footnote{\url{https://github.com/terrier-org/terrier-core/blob/5.x/doc/index.md}} to compute NDCG.

\subsection{Ranking Approaches} \label{sec:ranking_approaches}

We first include in our evaluation two baselines for reference and to allow comparing with the main ranking approaches considered. \\
\textbf{Random baseline.} The random baseline is included to provide a lower-bound of NDCG for the ranking task considered, and consists of randomly ranking products for each query. \\
\textbf{ESCI\_baseline.} The $ESCI\_baseline$ is the standard baseline introduced in the initial ESCI benchmark~\cite{esci}. It consists of using MS MARCO Cross-Encoders\footnote{\url{https://huggingface.co/cross-encoder/ms-marco-MiniLM-L-12-v2}} for the English subset, a Sentence Transformer model~\cite{sbert} trained on the MS Marco Passage Ranking task~\footnote{\url{https://github.com/microsoft/MSMARCO-Passage-Ranking}}. The model is further fine-tuned on the training set of SQD. The query and product title are used as input for the model.
\\
\\
The approaches evaluated in this paper and introduced below follow all the same core methodology for ranking: Cosine similarity is used to measure the relevance of a product to a query, and products are then ranked in decreasing order of similarity. The main difference lies in the models and data used to embed queries and products, used to compute similarity. \\
\textbf{SBERT\_text.} We use all-MiniLM-L12-v2\footnote{\url{https://huggingface.co/sentence-transformers/all-MiniLM-L12-v2}}, a Sentence Transformers model~\cite{sbert}, to embed queries and products. The query text and product title are used as input for the model.\\
\textbf{CLIP\_text.} We use CLIP~\cite{clip}, specifically clip-vit-large-patch14\footnote{\url{https://huggingface.co/openai/clip-vit-large-patch14}}, to embed queries and product titles. While the model is not specifically optimized for handling text alone, it enables the representation of text and images in the same space, which is required by some of the approaches considered here. \\
\textbf{CLIP\_image.} We use CLIP~\cite{clip}, specifically clip-vit-large-patch14, to embed queries and product images.
\\
\\
We also consider ranking approaches that combine both product text and images. This is done by either combining query-product similarities or directly combining ranking lists, using a weighted average. These approaches are designated by the notation ${m_{1}\_comb\_m_{2}}$, where $m_{1}$ and $m_{2}$ are the two approaches combined, and $comb$ is the method used to combine the results from $m_{1}$ and $m_{2}$ ($rank$ when combining rankings and $score$ when combining scores). A weight parameter $w$ is used to counterbalance the impact of text versus images. In terms of notation, $w$ is associated with $m_{1}$ and $(1-w)$ with $m_{2}$.

\section{Experimental Results} \label{sec:exp_results}

\begin{table*}[bp]
\caption{Experimental results}
\label{tab:exp_res_1}
\begin{center}
\vspace{-1.5em}
\begin{tabular}{|c|c|c|c|c|c|}
    \hline
    \textbf{Ranking approach} & \textbf{Random} & \textbf{ESCI\_Baseline} & \textbf{SBERT\_text} & \textbf{CLIP\_text} & \textbf{CLIP\_image} \\
    \hline
    \textbf{NDCG} & 0.7483 & 0.8562 & 0.8292 & 0.8107 & 0.8225 \\
    \hline
\end{tabular}
\end{center}
\vskip -0.2in
\end{table*}

Using the Terrier IR platform to compute NDCG for the $ESCI\_baseline$ leads to an NDCG of $0.83$, as reported in the ESCI benchmark~\cite{esci}. However, we noticed that the mapping of labels to relevance scores is interchanged for labels S and C in the code released (see line 48 in $prepare\_trec\_eval\_files.py$\footnote{\url{https://github.com/amazon-science/esci-data/blob/main/ranking/prepare_trec_eval_files.py}}). We thus adjusted the label-score mapping in the evaluation, leading to a different base NDCG score for the $ESCI\_baseline$. \\
\\
Table~\ref{tab:exp_res_1} presents the experimental results for the ranking approaches presented in Section~\ref{sec:ranking_approaches}. The random approach performs the worst and is only included for reference, to evaluate the impact of NDCG variations. As expected, the $ESCI\_baseline$ outperforms all the other approaches since it is the only method fine-tuned on the training data of SQD and leads to an NDCG of 0.8562. $SBERT\_text$ is followed very closely by $CLIP\_image$ and both perform better than $CLIP\_text$. The performance of $CLIP\_image$ shows the value of using images for the task of query-product ranking. \\

\begin{figure}[h]
  \centering
  \includegraphics[width=0.8\linewidth]{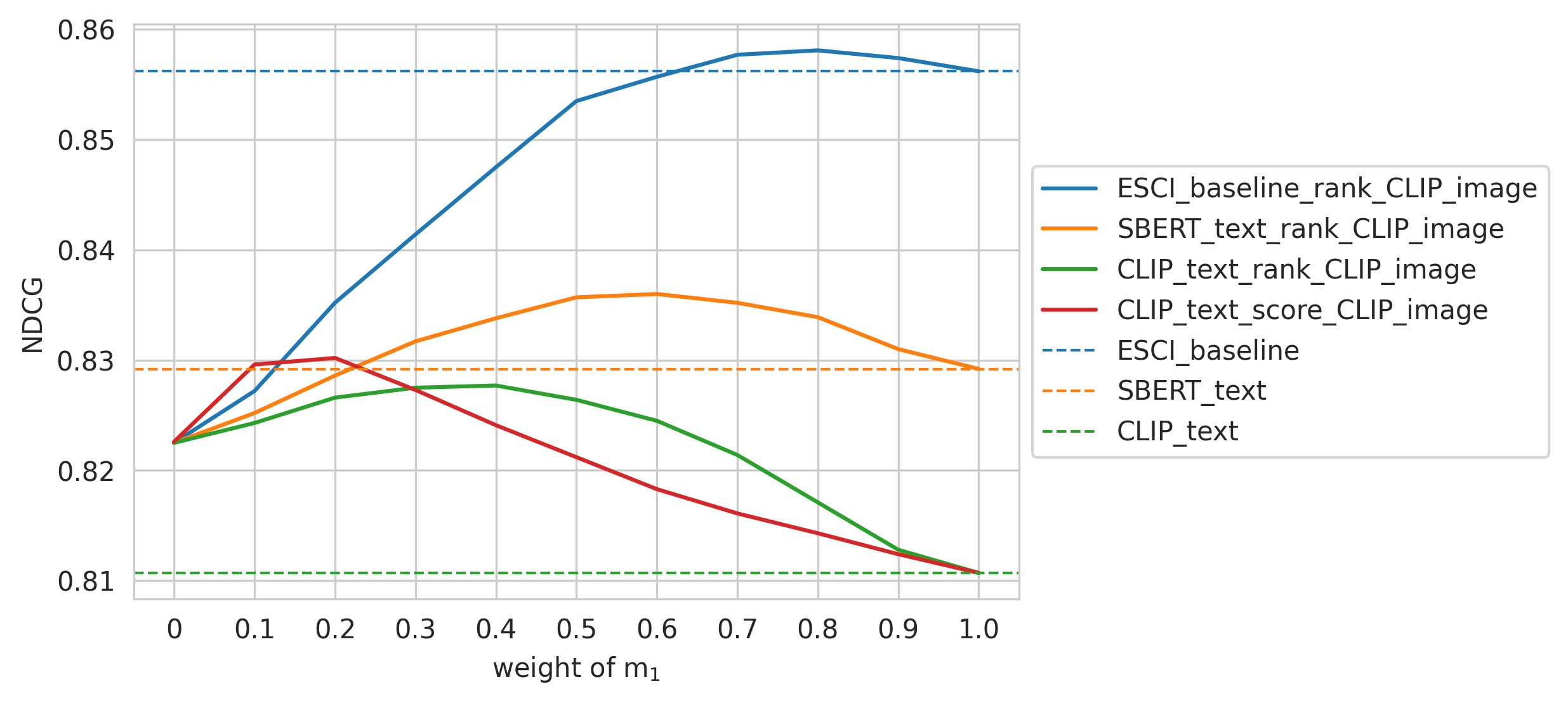}
  \caption{NDCG of combinations of ranking approaches, mixing both text and image data for query-product ranking. This is done by either combining query-product similarities or directly combining ranking lists, using a weighted average, where $w$ and $(1-w)$ are the weights associated respectively with $m_{1}$ and $m_{2}$ and $w$ is represented on the x-axis.}
  \label{fig:exp_res_2}
\end{figure}

Figure~\ref{fig:exp_res_2} shows the results for approaches combining both text and image. The performances of $ESCI\_baseline$, $SBERT\_text$, and $CLIP\_text$ are visualized as dashed horizontal lines on the plot, for reference. The points at $x=0.0$ correspond to the performance of $CLIP\_image$ (with a weight $w$ of $0.0$ for the text-based approach), and the points at $x=1.0$ correspond to the performance of the text-based approach (with a weight $(1 - w)$ of $0.0$ for the image-based approach). By varying the value of $w$, the weight of $m_{1}$, the results show that combining image and text outperforms the approach using only text data. \\

More specifically, $CLIP\_text\_score\_CLIP\_image$ reaches a lift of 2.41\% compared to the text-only approach (i.e., $CLIP\_text$),  $CLIP\_text\_rank\_CLIP\_image$ a lift of 2.1\%, $SBERT\_text\_rank\_CLIP\_image$ a lift of 0.82\%, and $ESCI\_baseline\_rank\_CLIP\_image$ a lift of 0.22\%. 

\section{Conclusion}

This paper presents the Shopping Queries Image Dataset (SQID), building upon the Amazon Shopping Dataset and enriching it with image information for products. We present the dataset and its characteristics, and provide experimental results showing the value of incorporating image data for the task of product search. We hope that this data will support further research around product search and ranking using multimodal data. 

SQID can be leveraged in the context of the ESCI benchmark, by evaluating the performance of models using images on the ESCI test set. The data can also be used to fine-tune pretrained models, outside of the ESCI benchmark. In addition, and as mentioned throughout the paper, the availability of text together with images allows investigating different techniques around multimodal learning relevant to the eCommerce space.

\begin{acknowledgments}
This dataset would not have been possible without the Shopping Queries Dataset by Amazon.
\end{acknowledgments}

\bibliography{sqid}

\end{document}